\documentclass{ws-ijmpa}

\newcommand{\be}{\begin{equation}}\newcommand{\ee}{\end{equation}}
\newcommand{\bea}{\begin{eqnarray}}\newcommand{\eea}{\end{eqnarray}}
\newcommand{\beaa}{\begin{eqnarray}}\newcommand{\eeaa}{\end{eqnarray}}
\newcommand{\ba}{\begin{array}}\newcommand{\ea}{\end{array}}
\newcommand{\bit}{\begin{itemize}}\newcommand{\eit}{\end{itemize}}
\newcommand{\ben}{\begin{enumerate}}\newcommand{\een}{\end{enumerate}}

\def\lab{\label}

\def\lf{\left}

\def\non{\nonumber}
\def\ran{\rangle}

\def\ri{\right}

\def\al{\alpha}\def\bt{\beta}
\def\De{\Delta}
\def\te{\theta}

\def\si{\sigma}
\def\om{\omega}

\def\1{{_{1}}}\def\2{{_{2}}}

\begin{document}

\markboth{M. Blasone, A. Capolupo, C.-R. Ji, G. Vitiello}
{On flavor conservation in weak interaction decays involving mixed neutrinos}

\catchline{}{}{}{}{}
\title{On flavor conservation in weak interaction decays  \\ involving mixed neutrinos}

\author{Massimo Blasone${}^{\flat}$, Antonio Capolupo${}^{\flat}$,
Chueng-Ryong Ji${}^{\sharp}$ and Giuseppe Vitiello${}^{\flat}$}
\address{
${}^{\flat}$ DMI,
 Universit\`a di Salerno and INFN,
 Gruppo Collegato di Salerno, 84100 Salerno, Italy,
 \\ ${}^{\sharp}$ Department of Physics, North Carolina State
University, Raleigh, NC 27695-8202, USA}

\maketitle

\pub{Received (Day Month Year)}{Revised (Day Month Year)}

\begin{abstract}

In the context of quantum field theory (QFT), we compute the amplitudes of weak
interaction processes such as $ W^{+}
\rightarrow e^{+} + \nu_{e}$ and $ W^{+} \rightarrow e^{+} +
\nu_{\mu}$ by using different
representations of flavor states for mixed neutrinos.
Analyzing the short time limit of the above amplitudes, we find that
the neutrino states defined in
QFT as eigenstates of the flavor charges lead to results consistent with
lepton charge conservation. On the contrary, the Pontecorvo flavor states produce a
violation of lepton charge in the vertex, which is in contrast with what
expected at tree
level in the Standard Model.

\end{abstract}

\ccode{PACS Nos.14.60.Pq, 13.38.Be, 03.70.+k}

\section{Introduction}

Given the importance of neutrino mixing and oscillations
\cite{Pontecorvo:1957cp} in elementary particle physics,
a great deal of work has been devoted to the related theoretical
issues. For example, in the definition of flavor states,
it has emerged \cite{BV95} that the vacuum for the mass
eigenstates of neutrinos turns out to be unitarily inequivalent to
the vacuum for the flavor eigenstates of neutrinos. The
vacuum structure associated with the field mixing
\cite{BV95,BHV98,currents,Fujii:1999xa,binger,JM01,yBCV02,Capolupo:2004av,Blasone:2005ae,Blasone:2005tk}
leads to a modification of the
flavor oscillation formulas \cite{BHV98,JM01,yBCV02} and
exhibits new features with respect to the  quantum mechanical
ones \cite{Pontecorvo:1957cp}. The theoretical understanding of the
mixing phenomena in the framework of the quantum field theory (QFT) has
also been confirmed by mathematically rigorous analysis
\cite{Hannabus}.  One of the offsprings from the QFT treatment
consists in the fact that it has led also  to consider, from the
perspective of particle mixing, other physically relevant problems
which would have not been possible to handle by resorting to the
Pontecorvo quantum mechanical (QM) approximation. For example, we
quote the particle
 mixing contribution to the dark energy of the Universe
\cite{Capolupo:2006et,Capolupo,Mavromatos}.

In this paper, we consider the concrete problem of verifying
the lepton flavor conservation in the processes that produce the (mixed)
neutrino.
Here, we analyze the amplitudes of weak interaction processes
such as $ W^{+} \rightarrow
e^{+} + \nu_{e}$ and $ W^{+} \rightarrow e^{+} + \nu_{\mu}$
at tree level in the context of QFT.
Although it is well known that the flavor changing loop
induced processes, such as $\mu \rightarrow e \gamma$, are possible,
they are not relevant to our discussion
since they have very low  branching ratios,
e.g.  $Br(\mu \rightarrow e \gamma)<10^{-50}$ \cite{Casas:2001sr}.

We carry out our calculations by resorting to two different
representations of flavor neutrino states:
1) the ones defined in Ref.\cite{BV95}, hereon denoted as
``exact flavor states'' and
2) the quantum mechanical  (Pontecorvo) flavor states.
In particular, we consider the  amplitudes in the short time range,
i.e. at very small distances from the production vertex.
We find that the use of the exact flavor states leads to  results consistent
with the lepton charge conservation as expected
in the Standard Model (SM) at tree level, whereas the Pontecorvo states
yield a violation of the lepton charge in the vertex.

Although obtained in different context, a similar violation has been found
in Ref.\cite{Nishi:2008sc}, where it has been shown that
the processes such as $\pi \rightarrow \mu \bar{\nu}_{e}$
are possible with a branching ratio much greater than the loop induced
processes as the one mentioned above.
The conclusion of Ref.\cite{Nishi:2008sc} was that an intrinsic flavor violation
for massive neutrinos would be present in the Standard
Model.
We show that such a violation arises as a consequence of an incorrect
choice for the (mixed) neutrino flavor states.

In Section II, we compute the
amplitudes of the weak interaction processes $ W^{+} \rightarrow
e^{+} + \nu_{e}$ and $ W^{+} \rightarrow e^{+} + \nu_{\mu}$ by using the exact flavor states
and the Pontecorvo states.
In Section III, we consider the explicit form of the above amplitudes for short time
intervals.
The long time limit is studied in Appendix B.
 Section \ref{S5} is devoted to conclusions. A brief summary of the vacuum structure
for  Dirac neutrino mixing is presented in the Appendix A.

\section{Amplitudes of weak interaction processes containing mixed neutrinos}

In this Section, we compute the amplitudes of the following two decays at tree
level:
\bea\label{We-decay} W^{+} \rightarrow e^{+} + \nu_{e}~,
\\[2mm]
\label{Wmu-decay} W^{+} \rightarrow e^{+} + \nu_{\mu}~,\eea
where neutrinos are produced through charged current processes.
Although our computations are specific for these decay processes,
our conclusions are general and hold for all the
different neutrino production processes.
We perform the calculations by means of standard QFT techniques.

In Section II.A, we use the exact flavor neutrino states
defined as the eigenstates of flavor charges
(see the Appendix for notations and definitions).
They are generated by the action of the flavor neutrino  creation operators on the flavor vacuum
$|0\ran_{f}$ as follows:
\bea\label{flavstate} |\nu^{r}_{\;{\bf k},\si} \ran \equiv
\al^{r\dag}_{{\bf k},{\nu_{\sigma}}} |0\ran_{f}, \qquad \si =
e,\mu \,, \eea
and $|\nu^{r}_{\;{\bf k},\sigma}(t) \ran = e^{ i H_{0} t}
\al^{r\dag}_{{\bf k},{\sigma}} |0\ran_{f} $.

In Section II.B,  we perform the same calculations of Section II.A by using the
quantum mechanical Pontecorvo states
\begin{eqnarray} \label{nue0a}
|\nu^{r}_{\;{\bf k},e}\rangle_P &=& \cos\theta\;|\nu^{r}_{\;{\bf
k},1}\rangle \;+\; \sin\theta\; |\nu^{r}_{\;{\bf k},2}\rangle \,,
\\ [2mm] \label{nue0b}
|\nu^{r}_{\;{\bf k},\mu}\rangle_P &=&
-\sin\theta\;|\nu^{r}_{\;{\bf k},1}\rangle \;+\; \cos\theta\;
|\nu^{r}_{\;{\bf k},2}\rangle \, ,
\end{eqnarray}
where the neutrino states with definite masses are defined
by the action of the creation operators for the free fields $\nu_i$
on the vacuum $|0\ran_{m}$ (see Appendix A):
\bea\label{massstate} |\nu^{r}_{\;{\bf k},i} \ran \equiv
\al^{r\dag}_{{\bf k},{i}} |0\ran_{m}, \qquad i =
1,2\,.
\eea
Note that the Pontecorvo neutrino states Eqs.(\ref{nue0a}),(\ref{nue0b}) are not eigenstates of
flavor neutrino charges \cite{Blasone:2005ae,Blasone:2005tk}.

In the scattering theory for finite range potentials, it is assumed \cite{Itzikson}
that the interaction Hamiltonian $H_{int}(x)$ can be
switched off adiabatically as $x^{0}_{in} \rightarrow -\infty$ and $x^{0}_{out}
\rightarrow +\infty$ so that the initial and final states can be
represented by the eigenstates of the free Hamiltonian.
However,
in the present case and more generally in the
decay processes where the mixed neutrinos are produced, the application
of the adiabatic hypothesis leads to erroneous conclusions (as made
in Ref.\cite{Li:2006qt}). Indeed, the flavor neutrino field
operators do not have the mathematical characterization necessary
to be defined as asymptotic field operators acting on the
massive neutrino vacuum. Moreover, the flavor states $|\nu_{{\bf
k},\sigma}^{r} \rangle$ are not eigenstates of the free Hamiltonian.
Therefore, the integration limits in the amplitudes of  decay processes
where mixed neutrinos are produced must be chosen so that the time
interval $ \Delta t = x^{0}_{out} - x^{0}_{in} $ is much shorter
than the characteristic neutrino oscillation time $t_{osc}$: $
\Delta t \ll t_{osc}$.

In this paper we consider at the first order of the perturbation theory
the amplitudes of the decays $(\ref{We-decay})$ and $(\ref{Wmu-decay})$.

In general, if $| \psi_{i} \rangle$ and $| \psi_{f} \rangle$ denote initial and final states,
the probability amplitude $ \langle \psi_{f}| e^{-i H t} | \psi_{i} \rangle$ is given by
\bea
\label{ref}\langle \psi_{f}| e^{-i H t} | \psi_{i} \rangle =
\langle e^{i H_{0} t}\psi_{f}| e^{i H_{0} t} e^{-i H t} | \psi_{i} \rangle =
\langle e^{i H_{0} t}\psi_{f}| U_{I}(t) | \psi_{i} \rangle\,.
\eea
Here the time evolution operator $U_{I}(t)$ in the interaction picture is given approximatively by
\bea
U_{I}(t) \simeq 1 - i \int_{0}^{t} dt' H_{int}(t')\,,
\eea
with $H_{int}(t) = e^{i H_{0} t} H_{int} e^{-i H_{0} t}$ interaction hamiltonian in the interaction picture.
In the following $H_0$ is the free part of the Hamiltonian for the fields involved in the decays
$(\ref{We-decay})$ and $(\ref{Wmu-decay})$ and the relevant interaction Hamiltonian  is given by \cite{chengli}:
 \bea \non
 H_{int}(x)
= - \frac{g}{\sqrt{2}}W_{\mu}^{+}(x)J_{W}^{\mu+}(x) + {\rm h.c.} = -
\frac{g}{2\sqrt{2}} W_{\mu}^{+}(x)
\overline{\nu}_{e}(x)\gamma^{\mu}(1-\gamma^{5})e(x)+ {\rm h.c.}, \\
 \label{H-interaction}\eea
where $W^{+}(x)$, $e(x)$ and $\nu_{e}(x)$ are the fields of the boson
$W^{+}$, the electron and the flavor (electron) neutrino,
respectively.

\subsection{Exact flavor states}

Let us first consider the process $W^{+} \rightarrow e^{+} + \nu_{e}~$
 and the states defined in Eq.(\ref{flavstate}).
The amplitude of the decay
at first order in perturbation theory is given by
\footnote{Note that in the case of the flavor states, because of the orthogonality of the Hilbert spaces
at different times (see Appendix A), instead of Eq.(\ref{ref}) the amplitude should be defined as
$\langle \psi_{\sigma}(x_{out}^0)|e^{-i H(x_{out}^0-x_{in}^0)}| \psi_{\sigma}(x_{in}^0) \rangle
=\langle \psi_{\sigma}(x^{0}_{in})|U_{I}(x_{out}^0, x_{in}^0)| \psi_{\sigma}(x^{0}_{in}) \rangle$.}

\bea\label{amplitude1}
 A_{W^{+} \rightarrow e^{+} + \nu_{e}} & = &  \langle \nu_{{\bf k},e}^{r},
e_{\bf q}^{s} | \lf[ -i \int_{x^{0}_{in}}^{x^{0}_{out}}  d^{4}x \,
H_{int}(x)  \ri] | W^{+}_{{\bf p},\lambda}\rangle
\\ \non & = &  _{W}\langle 0|\, \langle \nu_{{\bf k},e}^{r}(x^{0}_{in})|\, \langle e_{\bf q}^{s} |
\Big \{ \frac{i \,g }{2\sqrt{2}} \int_{x^{0}_{in}}^{x^{0}_{out}}
d^{4}x
\\ \non & \, & \Big[ W_{\mu}^{+}(x)
\overline{\nu}_{e}(x) \gamma^{\mu} (1-\gamma^{5}) e(x) \Big] \Big
\} | W^{+}_{{\bf p},\lambda}\rangle \,| 0 \rangle_{e} \, | 0(x^{0}_{in}) \rangle_{f}.
\eea
The terms involving the expectation values of the vector boson and electron fields
are given by
\bea
 \,_{W}\langle 0|W_{\mu}^{+}(x)| W^{+}_{{\bf p},\lambda}\rangle & = & \frac{1}{(2 \pi)^{3/2}}
\frac{\varepsilon _{{\bf p},\mu,\lambda}}{\sqrt{2
\omega_{p}}}\;e^{i({\bf p}\cdot{\bf x} - E^{W}_{p}x^{0})},
\\
 \langle e_{\bf q}^{s} |e(x)| 0 \rangle_{e} & = & \frac{1}{(2 \pi)^{3/2}}\;
v_{{\bf q},e}^{s} \;e^{-i({\bf q}\cdot{\bf x}-E^{e}_{q} x^{0})},
\eea
where $E^{W}_{p} = \sqrt{p^{2}+ M^{2}_{W}}$ and
$E^{e}_{q}= \sqrt{q^{2}+ M^{2}_{e}}$.

On the other hand, the term involving the expectation value of the
neutrino field yields
\bea  \label{nueexp}
&& \non\langle \nu_{{\bf k},e}^{r}(x_{in}^{0})|\overline{\nu}_{e}(x)| 0(x_{in}^{0})
\rangle_{f} \,= \, \frac{e^{-i{\bf k}\cdot{\bf
x}}}{(2 \pi)^{3/2}} \Big \{\overline{u}_{{\bf k},1}^{r}
\Big[\cos^{2}\theta \; e^{i \omega_{k,1} (x^{0}-x_{in}^{0})}
\\\non && + \sin^{2}\theta \Big( |U_{\bf k}|^{2}\; e^{i
\omega_{k,2} (x^{0}-x_{in}^{0})}
\, + \,   |V_{\bf k}|^{2} \; e^{-i
\omega_{k,2} (x^{0}-x_{in}^{0})} \Big) \Big]
 \\ && +
 \varepsilon^{r}\,|U_{\bf k}||V_{\bf k}|\, \overline{v}_{-{\bf k},1}^{r}
 \sin^{2}\theta
\Big[e^{-i \omega_{k,2}(x^{0}-x_{in}^{0})}-e^{i \omega_{k,2} (x^{0}-x_{in}^{0})} \Big]\Big \}\,,
 \eea
where
$ \omega _{k,i}=\sqrt{{\bf k}^{2} + m_{i}^{2}},$ $i=1,2$.
Here we have utilized the explicit form of the flavor annihilation/creation operators
given in Appendix A. Notice the presence in Eq.(\ref{nueexp}) of the Bogoliubov coefficients
$U_{\bf k}$ and $V_{\bf k}$.

It is also convenient to rewrite  Eq.(\ref{nueexp}),
by means of the relations (\ref{realzu2}) and
(\ref{realzv2}) given in Appendix A, as
\bea\non
&& \langle \nu_{{\bf k},e}^{r}(x_{in}^{0})|\overline{\nu}_{e}(x)| 0(x_{in}^{0})
\rangle_{f} \, = \,   \frac{e^{-i{\bf k}\cdot{\bf
x}}}{(2 \pi)^{3/2}}
 \;  \Big \{\cos^{2}\theta \; \overline{u}_{{\bf k},1}^{r}\; e^{i \omega_{k,1} (x^{0}-x_{in}^{0})}
\\ \label{nueexp2}
&& +
  \sin^{2}\theta \; \Big[\overline{u}_{{\bf k},2}^{r}\; |U_{\bf k}|\;e^{i \omega_{k,2} (x^{0}-x_{in}^{0})}
\;+\;
\varepsilon^{r}\; \overline{v}_{-{\bf k},2}^{r}\; |V_{\bf k}|\;
e^{-i \omega_{k,2} (x^{0}-x_{in}^{0})}\Big]\Big \}\;.
\eea

Combining the above results, Eq.(\ref{amplitude1}) can be written as
\bea\label{amplit}
 A_{W^{+} \rightarrow e^{+} + \nu_{e}}  & = &   \frac{i \,g }{2\sqrt{2}(2 \pi)^{3/2}}\;
\delta^{3}({\bf p}-{\bf q}-{\bf k})
\int_{x^{0}_{in}}^{x^{0}_{out}}  d x^{0} \; \frac{\varepsilon
_{{\bf p},\mu,\lambda}}{\sqrt{2 E^{W}_{p}}}\;
\\\non  & \times &
\Big \{
\overline{u}_{{\bf k},1}^{r} \gamma^{\mu}(1 - \gamma^{5}) v_{{\bf
q},e}^{s} \;
\Big [\cos^{2}\theta \;
 e^{- i (E^{W}_{p} - E^{e}_{q} - \omega_{k,1})x^{0}}\, e^{-i \omega_{k,1} x^{0}_{in}}
 \\ \non & + &
  \sin^{2}\theta \Big( |U_{\bf k}|^{2}
\; e^{- i (E^{W}_{p} - E^{e}_{q} - \omega_{k,2}) x^{0}} \, e^{-i \omega_{k,2} x^{0}_{in}}
\\ \non & + &  |V_{\bf k}|^{2}
\; e^{- i (E^{W}_{p} - E^{e}_{q} + \omega_{k,2}) x^{0}} \, e^{i \omega_{k,2} x^{0}_{in}} \Big)
\Big ]
\\ \non
& + & \varepsilon^{r}\,|U_{\bf k}||V_{\bf k}|\,
  \overline{v}_{-{\bf k},1}^{r} \gamma^{\mu}(1 - \gamma^{5}) v_{{\bf q},e}^{s}
  \sin^{2}\theta
 \\\non & \times &
\lf[e^{- i (E^{W}_{p} - E^{e}_{q} + \omega_{k,2}) x^{0}}\, e^{i \omega_{k,2} x^{0}_{in}} - e^{-
i (E^{W}_{p} - E^{e}_{q} - \omega_{k,2}) x^{0}}\, e^{- i \omega_{k,2} x^{0}_{in}} \ri]   \Big \},
\eea
when Eq.(\ref{nueexp}) is used, or equivalently as:
\bea\label{amplit2}\non
 A_{W^{+} \rightarrow e^{+} + \nu_{e}}  & = &   \frac{i \,g }{2\sqrt{2}(2 \pi)^{3/2}}\;
\delta^{3}({\bf p}-{\bf q}-{\bf k})
\int_{x^{0}_{in}}^{x^{0}_{out}}  d x^{0} \; \frac{\varepsilon
_{{\bf p},\mu,\lambda}}{\sqrt{2 E^{W}_{p}}}\;
\\ \non & \times &
\Big
\{\cos^{2}\theta\, e^{-i \omega_{k,1} x^{0}_{in}} \; \overline{u}_{{\bf k},1}^{r}\; \gamma^{\mu}(1
- \gamma^{5})\; v_{{\bf q},e}^{s} \; e^{-i(E^{W}_{p} -
E^{e}_{q} - \omega_{k,1})x^{0}} \nonumber
 \\ \non  & + &
  \sin^{2}\theta \; \Big[e^{-i \omega_{k,2} x^{0}_{in}}\;|U_{\bf k}|\; \overline{u}_{{\bf k},2}^{r}\,  \gamma^{\mu}(1 -
\gamma^{5})\; v_{{\bf q},e}^{s} \; e^{-i(E^{W}_{p} - E^{e}_{q}
- \omega_{k,2})x^{0}}
\\ & +&
e^{ i \omega_{k,2} x^{0}_{in}}\;\varepsilon^{r}\, |V_{\bf k}|\; \overline{v}_{-{\bf k},2}^{r}\,
\gamma^{\mu}(1 - \gamma^{5})\; v_{{\bf q},e}^{s} \;
e^{-i(E^{W}_{p} - E^{e}_{q} + \omega_{k,2})x^{0}}\Big]\Big \}\;,
\eea
when the term involving the expectation value of the
neutrino field is expressed in the form of Eq.(\ref{nueexp2}).

\vspace{0.5cm}

Next we consider the process $ W^{+} \rightarrow e^{+} + \nu_{\mu}$.
By using the  Hamiltonian  (\ref{H-interaction}), we have now
\bea\label{amplitude2}
 A_{W^{+} \rightarrow e^{+} + \nu_{\mu}}  &= & \langle \nu_{{\bf k},\mu}^{r},
e_{\bf q}^{s} | \lf[ -i \int_{x^{0}_{in}}^{x^{0}_{out}}  d^{4}x \,
H_{int}(x)  \ri] | W^{+}_{{\bf p},\lambda}\rangle
\\  \nonumber
& =&  \,_{W}\langle 0|\, \langle \nu_{{\bf k},\mu}^{r}(x^{0}_{in})|\, \langle e_{\bf q}^{s} |
\Big \{ \frac{i \,g }{2\sqrt{2}}
\\  \nonumber
& \times & \int_{x^{0}_{in}}^{x^{0}_{out}}
d^{4}x
 \Big[ W_{\mu}^{+}(x)
\overline{\nu}_{e}(x)\gamma^{\mu}(1-\gamma^{5})e(x) \Big] \Big \}
\,| W^{+}_{{\bf p},\lambda}\rangle \,| 0 \rangle_{e} \, | 0(x^{0}_{in})
\rangle_{f}.
\eea

The term involving the expectation value of the
neutrino field is now
\bea\label{numuexp}\nonumber
 && \langle \nu_{{\bf k},\mu}^{r}(x^{0}_{in})|\overline{\nu}_{e}(x)| 0(x^{0}_{in})
\rangle_{f} \, = \, \frac{e^{-i{\bf k}\cdot{\bf
x}}}{(2 \pi)^{3/2}}\sin \theta \cos \theta  \Big \{|U_{\bf k}| \;
\overline{u}_{{\bf k},1}^{r}\; \Big[  e^{i \omega_{k,2} (x^{0}-x^{0}_{in})}
\\ && -  e^{i \omega_{k,1} (x^{0}-x^{0}_{in})}\Big]
 +
 \varepsilon^{r}\,|V_{\bf k}|\, \overline{v}_{-{\bf k},1}^{r}
\lf[\;- e^{i \omega_{k,2} (x^{0}-x^{0}_{in})} +  e^{-i
\omega_{k,1} (x^{0}-x^{0}_{in})}\ri]\Big \},
\eea
which, using the relations (\ref{realzu2}) and
(\ref{realzv2}) in Appendix A,  can be also written as
\bea\non\label{numuexp2}
& & \langle \nu_{{\bf k},\mu}^{r}(x^{0}_{in})|\overline{\nu}_{e}(x)| 0(x^{0}_{in})
\rangle_{f} \,= \, \frac{e^{-i{\bf k}\cdot{\bf
x}}}{(2 \pi)^{3/2}}\sin \theta \cos \theta  \Big \{
\overline{u}_{{\bf k},2}^{r}\; e^{i \omega_{k,2} (x^{0}-x^{0}_{in})}
\\ &&- |U_{\bf k}| \;
\overline{u}_{{\bf k},1}^{r}\;  e^{i \omega_{k,1} (x^{0}-x^{0}_{in})}
 +
 \varepsilon^{r}\,|V_{\bf k}|\, \overline{v}_{-{\bf k},1}^{r}
 e^{-i \omega_{k,1}(x^{0}-x^{0}_{in})}\Big \}.
\eea
Thus, the amplitude (Eq.(\ref{amplitude2})) can be expressed as
\bea
\non\label{amplitude2expl}
 A_{W^{+} \rightarrow e^{+} + \nu_{\mu}} &=&   \frac{i \,g }{2\sqrt{2}(2 \pi)^{3/2}}\;
 \delta^{3}({\bf p}-{\bf q}-{\bf k})\;\sin \theta \cos \theta \;
\int_{x^{0}_{in}}^{x^{0}_{out}}  d x^{0} \; \frac{\varepsilon
_{{\bf p},\mu,\lambda}}{\sqrt{2 E^{W}_{p}}}\;
\\\non & \times &
 \Big \{
|U_{\bf k}|\,\overline{u}_{{\bf k},1}^{r} \gamma^{\mu}(1 -
\gamma^{5}) v_{{\bf q},e}^{s} \;
 \Big[\; e^{-i (E^{W}_{p} - E^{e}_{q} - \omega_{k,2})x^{0}}\, e^{-i \omega_{k,2} x^{0}_{in}}
 \\\non &  - &  e^{-i (E^{W}_{p} - E^{e}_{q} - \omega_{k,1})x^{0}}\, e^{-i \omega_{k,1} x^{0}_{in}}\Big]
 \\\non &  +&
  \varepsilon^{r}\; |V_{\bf k}|\,
 \overline{v}_{-{\bf k},1}^{r} \gamma^{\mu}(1 - \gamma^{5}) v_{{\bf q},e}^{s}
\Big[-e^{- i (E^{W}_{p} - E^{e}_{q} - \omega_{k,2}) x^{0}}\, e^{-i \omega_{k,2} x^{0}_{in}}
\\ &  + & e^{- i
(E^{W}_{p} - E^{e}_{q} + \omega_{k,1}) x^{0}}\, e^{ i \omega_{k,1} x^{0}_{in}} \Big] \Big \}\;,
\eea
when Eq.(\ref{numuexp}) is utilized, or equivalently as
\bea\label{amplitude2expl2}\non
 A_{W^{+} \rightarrow e^{+} +
\nu_{\mu}} & = &   \frac{i \,g }{2\sqrt{2}(2 \pi)^{3/2}}\;
 \delta^{3}({\bf p}-{\bf q}-{\bf k})\;\sin \theta \cos \theta \;
\int_{x^{0}_{in}}^{x^{0}_{out}}  d x^{0} \; \frac{\varepsilon
_{{\bf p},\mu,\lambda}}{\sqrt{2 E^{W}_{p}}}\;
\\ & \times &
\Big [e^{- i \omega_{k,2} x^{0}_{in}}\;
\overline{u}_{{\bf k},2}^{r}\,  \gamma^{\mu}(1 - \gamma^{5})\;
v_{{\bf q},e}^{s} \; e^{-i(E^{W}_{p} - E^{e}_{q} -
\omega_{k,2})} \nonumber
 \\  & - & \non
  \;e^{- i \omega_{k,1} x^{0}_{in}}
  \; |U_{\bf k}|\;\overline{u}_{{\bf k},1}^{r}\,  \gamma^{\mu}(1 - \gamma^{5})\; v_{{\bf q},e}^{s} \;
e^{-i(E^{W}_{p} - E^{e}_{q} - \omega_{k,1})}
\\ & + & \, e^{ i \omega_{k,1} x^{0}_{in}}
\varepsilon^{r}\,|V_{\bf k}|\; \overline{v}_{-{\bf k},1}^{r}
\gamma^{\mu}(1 - \gamma^{5}) v_{{\bf q},e}^{s} \;e^{-i(E^{W}_{p}
- E^{e}_{q} + \omega_{k,1})} \Big ]\;,
\eea
when Eq.(\ref{numuexp2}) is used.

\subsection{Pontecorvo flavor states\label{PontAmpl}}

We now repeat the computations using the Pontecorvo
states (\ref{nue0a}) and (\ref{nue0b}) instead of the exact flavor states
(\ref{flavstate}).

For the decay  $ W^{+} \rightarrow e^{+} + \nu_{e}$,
we have (the superscript $P$ denotes the amplitude computed with Pontecorvo states)
\bea\label{amplitude1Pont}
 A^P_{W^{+} \rightarrow e^{+} + \nu_{e}} & = &
  _{W}\langle 0|\, _{P}\langle \nu_{{\bf k},e}^{r}(x^{0}_{out})|\, \langle e_{\bf q}^{s} |
\Big \{ \frac{i \,g }{2\sqrt{2}}
\\\non & \times &
\int_{x^{0}_{in}}^{x^{0}_{out}}
d^{4}x \, \lf[ W_{\mu}^{+}(x)\;
\overline{\nu}_{e}(x)\;\gamma^{\mu}\;(1-\gamma^{5})\;e(x) \ri] \Big
\} | W^{+}_{{\bf p},\lambda}\rangle \,| 0 \rangle_{e} \, | 0
\rangle_{m}.
\eea
In the above expressions, notice that the vacuum $| 0 \rangle_{m}$ appears
for the fields with definite masses $\nu_{i}$.

In the amplitude (\ref{amplitude1Pont}),
the term involving the expectation value of the neutrino fields given in  Eq.(\ref{nueexp})
(or equivalently in Eq.(\ref{nueexp2})) is replaced by
 \bea
&& \;_{P}\langle \nu_{{\bf k},e}^{r}(x^{0}_{out})|\overline{\nu}_{e}(x)| 0
\rangle_{m} \,=\,
\\\non &&  =  \cos \theta\;e^{-i  \omega_{k,1} x^{0}_{out}} \langle \nu_{{\bf
k},1}^{r}|\;\overline{\nu}_{e}(x)\;| 0 \rangle_{m}
\, + \, \sin \theta\;e^{-i  \omega_{k,2} x^{0}_{out}} \langle \nu_{{\bf k},2}^{r}|\;\overline{\nu}_{e}(x)\;| 0
\rangle_{m}
\\\non \label{nueponte1}
&& =
\frac{e^{-i {\bf k}\cdot {\bf x}}}{(2 \pi)^{3/2}}\; \Big[\cos^{2}\theta\;\overline{u}_{{\bf
k},1}^{r}\; e^{-i  \omega_{k,1}( x^{0}_{out} - x^{0})}\;
 + \; \sin^{2}\theta\; \overline{u}_{{\bf
k},2}^{r}\; e^{ -i  \omega_{k,2}( x^{0}_{out} - x^{0})}\Big],
\eea
with respect to the amplitude (\ref{amplitude1}) computed with the exact flavor
states.
Thus, the amplitude $A^{P}_{W^{+} \rightarrow e^{+} + \nu_{e}}$
 becomes
\bea\label{amplitude-nuePonte1}\non
 A^{P}_{W^{+} \rightarrow e^{+} + \nu_{e}} & = &  \frac{i \,g }{2\sqrt{2}(2 \pi)^{1/2}}
 \; \frac{\varepsilon _{{\bf p},\mu,\lambda}}{\sqrt{2 E^{W}_{p}}}\;\delta^{3}({\bf p}-{\bf q}-{\bf k})\;
\\ \non & \times &
\int_{x^{0}_{in}}^{x^{0}_{out}}  d x^{0} \; \Big [\cos^{2}\theta \;e^{- i  \omega_{k,1} x^{0}_{out}}\;
\overline{u}_{{\bf k},1}^{r}\; \gamma^{\mu}(1 - \gamma^{5})\;
v_{{\bf q},e}^{s} \; e^{-i (E^{W}_{p} - E^{e}_{q} - \omega_{k,1})x^{0}}
  \nonumber
 \\  & + & \;
  \sin^{2}\theta \;e^{- i  \omega_{k,2} x^{0}_{out}}\; \overline{u}_{{\bf k},2}^{r}\;
  \gamma^{\mu}(1 - \gamma^{5})\; v_{{\bf q},e}^{s} \; e^{-i (E^{W}_{p} - E^{e}_{q} - \omega_{k,2})x^{0}}
\Big ]\;.
\eea

\vspace{0.5cm}

In a similar way, when we consider the decay  $ W^{+} \rightarrow e^{+} +
\nu_{\mu}$, the expectation value given in Eq.(\ref{numuexp})
(or equivalently in Eq.(\ref{numuexp2})) is replaced by
 \bea
&& \;_{P}\langle \nu_{{\bf k},\mu}^{r}(x^{0}_{out})|\overline{\nu}_{e}(x)| 0 \rangle_{m}  =
\\\non
&& - \sin \theta\;e^{- i  \omega_{k,1} x^{0}_{out}}\; \langle \nu_{{\bf
k},1}^{r}|\;\overline{\nu}_{e}(x)\;| 0 \rangle_{m}
\,+ \,  \cos \theta\;e^{- i  \omega_{k,2} x^{0}_{out}}\; \langle \nu_{{\bf k},2}^{r}|\;\overline{\nu}_{e}(x)\;| 0 \rangle_{m}
\\ \non\label{nueponte2}
 && =
\frac{ e^{-i {\bf k}\cdot {\bf x}} \sin \theta \cos \theta}{(2 \pi)^{3/2}}\; \lf[
\overline{u}_{{\bf k},2}^{r}\; e^{-i \omega_{k,2}( x^{0}_{out}-x^{0})}\; - \;
 \overline{u}_{{\bf k},1}^{r}\;
e^{-i  \omega_{k,1}( x^{0}_{out}-x^{0})} \ri],
\eea
and the amplitude $A^{P}_{W^{+} \rightarrow e^{+} + \nu_{\mu}}$ is now given by
\bea\label{amplitude-numuPonte1}\non
  A^{P}_{W^{+} \rightarrow e^{+} + \nu_{\mu}} & = & \frac{i \,g }{2\sqrt{2}(2 \pi)^{1/2}}
 \; \frac{\varepsilon _{{\bf p},\mu,\lambda}}{\sqrt{2 E^{W}_{p}}}\; \sin \theta\; \cos
\theta\,\delta^{3}({\bf p}-{\bf q}-{\bf k})\,
\\ \non & \times &
\int_{x^{0}_{in}}^{x^{0}_{out}}  d x^{0} \; \Big [e^{- i  \omega_{k,2} x^{0}_{out}}\;
\overline{u}_{{\bf k},2}^{r}\; \gamma^{\mu}\; (1 - \gamma^{5})\;
v_{{\bf q},e}^{s} \; \; e^{-i (E^{W}_{p} - E^{e}_{q} - \omega_{k,2})x^{0}}
\; \\ & - &
 \;e^{- i  \omega_{k,1} x^{0}_{out}}\; \overline{u}_{{\bf k},1}^{r}\;
  \gamma^{\mu}\; (1 - \gamma^{5})\; v_{{\bf q},e}^{s} \;
 \; e^{-i (E^{W}_{p} - E^{e}_{q} - \omega_{k,1})x^{0}}\Big ]\;.
\eea
In the relativistic limit,
$|V_{\bf k}| \rightarrow 0$ and $|U_{\bf k}| \rightarrow
1$, and, regardless phase factors, Eqs.(\ref{amplit2}) and (\ref{amplitude2expl2}) coincide
with  Eqs.(\ref{amplitude-nuePonte1}) and (\ref{amplitude-numuPonte1}),
respectively, obtained by using the Pontecorvo states.
These are
 indeed the approximation of the exact flavor states
in such a relativistic limit \cite{BV95}.

The general expressions given by Eqs.(\ref{amplit2}) and (\ref{amplitude2expl2})
as well as those given by Eqs.(\ref{amplitude-nuePonte1}) and (\ref{amplitude-numuPonte1}) will be
the basis for our analysis of lepton charge conservation for the processes
$ W^{+} \rightarrow e^{+} + \nu_{e}$ and $ W^{+} \rightarrow e^{+} + \nu_{\mu}$.
For this purpose, in next Section, we study the detailed structure of the  amplitudes of these processes
in the short time limit.

In the Appendix B, we also comment on the above amplitudes in the long time
limit.

\section{Amplitudes in the short time limit}

In this Section we consider the explicit form of the amplitudes given by
Eqs.(\ref{amplit2}), (\ref{amplitude2expl2}),
 (\ref{amplitude-nuePonte1}) and (\ref{amplitude-numuPonte1}) for short time intervals $\De t$.
The physical meaning of such a time scale $\De t$ is represented by the relation
 $\frac{1}{\Gamma}\ll \De t \ll L_{osc}$, where
$\Gamma$ is the $W^{+}$ decay width and  $L_{osc}$ is the typical flavor oscillation length.
Given the experimental values of $\Gamma$ and $L_{osc}$,
this interval is well defined. A similar assumption
has been made in Ref.\cite{Nishi:2008sc}  where the decay
$\pi \rightarrow \mu \bar{\nu}_{e}$ has been considered.
In the following, when we use the expression ``short time limit'', we refer
to the time scale defined above. Of course,  energy fluctuations are constrained by the Heisenberg uncertainty relation, where $\De t$ is the one given above.

We will see that the use of the exact flavor states gives results
which agree with lepton charge conservation in the production vertex,
as predicted (at tree level) by the SM.
On the other hand, we observe a clear violation of the lepton charge
when the Pontecorvo states are used.
Our calculation shows that the origin of such a violation is due to the
fact that the Pontecorvo flavor states are defined by use of the vacuum state $|0 \rangle_m$
for the massive neutrino states.
%

\subsection{Exact flavor states}

Let us first  consider the decay $W^{+} \rightarrow e^{+} + \nu_{e}$.
Taking the limit of integrations in Eq.(\ref{amplit2}) as
$x^{0}_{in} = -\De t/2$ and $x^{0}_{out} = \De t/2$, the amplitude
$A_{W^{+} \rightarrow e^{+} + \nu_{e}}$ becomes
\bea\label{amplitude-nue4}
&& A_{W^{+} \rightarrow e^{+} + \nu_{e}}  \, = \,  \frac{i \,g }{\sqrt{2}(2 \pi)^{3/2}}
 \; \frac{\varepsilon _{{\bf p},\mu,\lambda}}{\sqrt{2 E^{W}_{p}}}\;\delta^{3}({\bf p}-{\bf q}-{\bf k})\;
\\\non && \times
 \Big\{\cos^{2}\theta \;e^{ i \omega_{k,1} \De t/2} \; \overline{u}_{{\bf k},1}^{r}\, \frac{\sin[(E^{W}_{p} -
E^{e}_{q} - \omega_{k,1}) \De t/2]}{E^{W}_{p} -
E^{e}_{q} - \omega_{k,1}} 
 \\ \non  && +
 \sin^{2}\theta \; \Big[e^{ i \omega_{k,2} \De t/2}\;|U_{\bf k}|\; \overline{u}_{{\bf k},2}^{r}\,
   \frac{\sin[(E^{W}_{p} - E^{e}_{q}- \omega_{k,2}) \De t/2]}{E^{W}_{p} - E^{e}_{q}
- \omega_{k,2}}
 \\ \non  && + \,e^{ -i \omega_{k,2} \De t/2}\;
 \varepsilon^{r}\, |V_{\bf k}|\; \overline{v}_{-{\bf k},2}^{r}\,
\frac{\sin[(E^{W}_{p} - E^{e}_{q} + \omega_{k,2}) \De t/2]}{E^{W}_{p} - E^{e}_{q} + \omega_{k,2}}\Big]\Big\}\;
\; \gamma^{\mu}(1
- \gamma^{5})\; v_{{\bf q},e}^{s}.
\eea

We now consider the short time limit
of the above expression. It is clear
that the dominant contributions in Eq.(\ref{amplitude-nue4})
are those for which $E^{W}_{p} - E^{e}_{q} \mp \omega_{k,i} \approx 0$.
For such dominant terms, it is then safe to perform the expansion $\sin x \simeq x$.
Moreover we performe the expansion $e^{\pm i \omega_{k,i} \De t/2} \simeq 1$, with $i=1,2$. We thus  obtain
the following result at first order in $\De t$:
\bea\label{amplitude-nue5}
 A_{W^{+} \rightarrow e^{+} + \nu_{e}} & \simeq &  \frac{i \,g }{2\sqrt{2}(2 \pi)^{3/2}}
 \; \frac{\varepsilon _{{\bf p},\mu,\lambda}}{\sqrt{2 E^{W}_{p}}}\;\delta^{3}({\bf p}-{\bf q}-{\bf k})\; \De t\;\times
\\ \non &\times & \Big
\{\cos^{2}\theta \; \overline{u}_{{\bf k},1}^{r}
 +   \sin^{2}\theta \; \Big[|U_{\bf k}|\; \overline{u}_{{\bf k},2}^{r}
+ \varepsilon^{r}\, |V_{\bf k}|\; \overline{v}_{-{\bf k},2}^{r}\Big]\Big \}\; \gamma^{\mu}(1 - \gamma^{5})\; v_{{\bf q},e}^{s}.
\eea

The quantity in the curly brackets can be evaluated by means of the identity
given by Eq.(\ref{realzu3}) among the Bogoliubov coefficients.
The result is
\bea\label{amplitude-nue6}
 A_{W^{+} \rightarrow e^{+} + \nu_{e}} & \simeq &  \frac{i \,g }{2\sqrt{2}(2 \pi)^{3/2}}
 \; \frac{\varepsilon _{{\bf p},\mu,\lambda}}{\sqrt{2 E^{W}_{p}}}\;
 \delta^{3}({\bf p}-{\bf q}-{\bf k})\;\De t\;  \overline{u}_{{\bf k},1}^{r}
 \; \gamma^{\mu}(1 - \gamma^{5})\; v_{{\bf q},e}^{s}\,.
\eea
This amplitude resembles the one for the production of a free neutrino with mass $m_{1}$.

\medskip

Let us now turn to the process $W^{+} \rightarrow e^{+} + \nu_{\mu}$.
Proceeding in a similar way as above, taking
$x^{0}_{in} = -\De t/2$ and $x^{0}_{out} = \De t/2$ in Eq.(\ref{amplitude2expl2}),
we get
\bea\label{amplitude-numuFinal2} \non
&& A_{W^{+} \rightarrow e^{+} +
\nu_{\mu}} \, = \,  \frac{i \,g }{ 2\sqrt{2}(2 \pi)^{3/2}}
 \; \frac{\varepsilon _{{\bf p},\mu,\lambda}}{\sqrt{2 E^{W}_{p}}}\; \delta^{3}({\bf p}-{\bf
q}-{\bf k})\;
 \sin 2\theta \;
 \\ \non && \times
\Big [e^{  i \omega_{k,2} \De t/2}\;\overline{u}_{{\bf k},2}^{r}\,  \frac{ \sin [(E^{W}_{p} - E^{e}_{q}
- \omega_{k,2}) \De t/2]}{E^{W}_{p} - E^{e}_{q} - \omega_{k,2}}
\\ \non && -
e^{  i \omega_{k,1} \De t/2}\; |U_{\bf k}|\;\overline{u}_{{\bf k},1}^{r}\,
\frac{\sin [(E^{W}_{p} - E^{e}_{q} - \omega_{k,1}) \De t/2]}{E^{W}_{p} - E^{e}_{q} - \omega_{k,1}}
\\  && +
\, e^{ - i \omega_{k,2} \De t/2} \varepsilon^{r}\,|V_{\bf k}|\;\overline{v}_{-{\bf k},1}^{r}\;
\frac{\sin [(E^{W}_{p} - E^{e}_{q} + \omega_{k,1}) \De t/2]}{E^{W}_{p} - E^{e}_{q} + \omega_{k,1}} \Big ]
\; \gamma^{\mu}(1 - \gamma^{5})\;
v_{{\bf q},e}^{s}\,,
\eea
which becomes
\bea\label{amplitude-numuFinal3}\non
 \non A_{W^{+} \rightarrow e^{+} +
\nu_{\mu}} & \simeq &  \frac{i \,g }{4\sqrt{2}(2 \pi)^{3/2}}
 \; \frac{\varepsilon _{{\bf p},\mu,\lambda}}{\sqrt{2 E^{W}_{p}}}\; \delta^{3}({\bf p}-{\bf
q}-{\bf k})\; \De t\; \sin 2 \theta
\\ & \times &
\Big [
\overline{u}_{{\bf k},2}^{r}
 \;-\;
|U_{\bf k}|\;\overline{u}_{{\bf k},1}^{r}
\, +  \,\varepsilon^{r}\,|V_{\bf k}|\; \overline{v}_{-{\bf k},1}^{r}
 \Big ]\; \gamma^{\mu}(1 - \gamma^{5})\;
v_{{\bf q},e}^{s}\,,
\eea
in the short time limit.

We now observe that the quantity in square bracket vanishes identically due to
the relation given by Eq.(\ref{realzu2}): i.e.
\bea\label{amplitude-numuFinal4}
A_{W^{+} \rightarrow e^{+} +
\nu_{\mu}} \simeq 0\,.
\eea

This proves that, in the short time limit, the use of the exact flavor states leads
to the conservation of lepton charge in the production vertex in agreement with
what we expected
from the Standard Model.

\subsection{Pontecorvo states}

It is now straightforward to analyze the short time limit of the amplitudes
$A^{P}_{W^{+} \rightarrow e^{+} +
\nu_{e}}$ and $A^{P}_{W^{+} \rightarrow e^{+} +
\nu_{\mu}}$ defined by means of the Pontecorvo flavor states.

Proceeding in the same way as done in the previous subsection,
 Eq.(\ref{amplitude-nuePonte1})
becomes
 \bea\label{amplitude-nuePontec}\non
 A^{P}_{W^{+} \rightarrow e^{+} + \nu_{e}} & \simeq &  \frac{i \,g }{2\sqrt{2}(2 \pi)^{3/2}}
 \; \frac{\varepsilon _{{\bf p},\mu,\lambda}}{\sqrt{2 E^{W}_{p}}}\;\delta^{3}({\bf p}-{\bf q}-{\bf k})\;\De t
\\
& \times &
 \lf[\cos^{2}\theta \;\overline{u}_{{\bf k},1}^{r} + \sin^{2}\theta \;\overline{u}_{{\bf k},2}^{r}\ri ]
 \; \gamma^{\mu}(1 - \gamma^{5})\; v_{{\bf q},e}^{s}\;,
\eea
where we performed the expansion $e^{-i \omega_{k,i}\De t/2} \simeq 1$, with $i=1,2$.
The structure of this amplitude is clearly different from the one obtained in Eq.(\ref{amplitude-nue6}).
Such a difference is more relevant in the non-relativistic limit.

However, observed neutrinos are relativistic and thus is convenient to consider the relativistic limit
of the above result. To this end, we rewrite Eq.(\ref{amplitude-nuePontec})
in a more convenient form by using the identity given by Eq.(\ref{realzu2}):
\bea\label{amplitude-nuePontec1}
 A^{P}_{W^{+} \rightarrow e^{+} + \nu_{e}} & \simeq &  \frac{i \,g }{2\sqrt{2}(2 \pi)^{3/2}}
 \; \frac{\varepsilon _{{\bf p},\mu,\lambda}}{\sqrt{2 E^{W}_{p}}}\;\delta^{3}({\bf p}-{\bf q}-{\bf k})\;\De t
\\ \non & \times &
 \;  \Big[\overline{u}_{{\bf k},1}^{r}\lf(1 -  \sin^{2}\theta \; (1-|U_{\bf k}|)\ri)\, -
  \sin^{2}\theta \;\varepsilon^{r}\, \overline{v}_{-{\bf k},1}^{r}\,|V_{\bf k}|\Big]
 \; \gamma^{\mu}(1 - \gamma^{5})\; v_{{\bf q},e}^{s}.
 \eea
In the relativistic limit, the Bogoliubov coefficient $|U_{\bf k}| $ and
$|V_{\bf k}|$ can be expressed respectively as (see Appendix A):
\bea\label{relU-V}
|U_{\bf k}| \sim 1- \frac{(\Delta m)^2}{4 k^2}\,,\qquad \qquad|V_{\bf k}| \sim \frac{\Delta m}{2 k}\,,
\eea
where $\Delta m = m_2 -m_1$. Eq.(\ref{amplitude-nuePontec1}) can be then written,
at the first order in  $O\lf(\frac{\Delta m}{2 k}\ri)$,
 as
\bea\label{amplitude-nuePontec2}\non
 A^{P}_{W^{+} \rightarrow e^{+} + \nu_{e}} & \simeq &  \frac{i \,g }{2\sqrt{2}(2 \pi)^{3/2}}
 \; \frac{\varepsilon _{{\bf p},\mu,\lambda}}{\sqrt{2 E^{W}_{p}}}\;\delta^{3}({\bf p}-{\bf q}-{\bf k})\;\De t
 \\
 & \times &
   \lf[\overline{u}_{{\bf k},1}^{r} -
  \sin^{2}\theta \;\varepsilon^{r}\,  \overline{v}_{-{\bf k},1}^{r}\,\frac{\Delta m}{2 k}\ri ]
 \; \gamma^{\mu}(1 - \gamma^{5})\; v_{{\bf q},e}^{s}\,,
\eea
which shows how the results (\ref{amplitude-nue6}) and (\ref{amplitude-nuePontec2})
 agree in the ultra-relativistic limit (i.e. when $\frac{\Delta m}{ k} \rightarrow 0$).

\medskip

We now consider the short time limit of the amplitude given in Eq.(\ref{amplitude-numuPonte1}).
We have
\bea\label{amplitude-numuPontec3}\non
 A^{P}_{W^{+} \rightarrow e^{+} +
\nu_{\mu}} & \simeq &  \frac{i \,g }{4\sqrt{2}(2 \pi)^{3/2}}
 \; \frac{\varepsilon _{{\bf p},\mu,\lambda}}{\sqrt{2 E^{W}_{p}}}\; \delta^{3}({\bf p}-{\bf
q}-{\bf k})\; \sin 2\theta  \; \De t
\\ & \times &
 \Big [
\overline{u}_{{\bf k},2}^{r} -
 \overline{u}_{{\bf k},1}^{r}
\Big ]
\; \gamma^{\mu}(1 - \gamma^{5})\; v_{{\bf q},e}^{s}\,,
\eea
which signals a violation of lepton charge in the tree level vertex.
We performed the expansion $e^{-i \omega_{k,i}\De t/2} \simeq 1$, with $i=1,2$.

Again, we consider the relativistic limit. We first rewrite Eq.(\ref{amplitude-numuPontec3})
by means of the relation given by Eq.(\ref{realzu2}),
\bea\label{amplitude-numuPontec4}\non
{}\hspace{-1cm} A^{P}_{W^{+} \rightarrow e^{+} +
\nu_{\mu}} & \simeq &  \frac{i \,g }{4\sqrt{2}(2 \pi)^{3/2}}
 \; \frac{\varepsilon _{{\bf p},\mu,\lambda}}{\sqrt{2 E^{W}_{p}}}\; \delta^{3}({\bf p}-{\bf
q}-{\bf k})\; \sin 2\theta  \; \De t
\\ & \times &
\Big [\overline{u}_{{\bf k},1}^{r}(|U_{\bf k}|-1)
 -  \varepsilon^{r}\, \overline{v}_{-{\bf k},1}^{r}|V_{\bf k}|
\Big ]
\; \gamma^{\mu}(1 - \gamma^{5})\; v_{{\bf q},e}^{s}\,,
\eea
and, by using Eq.(\ref{relU-V}),
we obtain the following result at first order in $\frac{\Delta m}{2 k}$:
\bea\label{amplitude-numuPontec5}\non
 A^{P}_{W^{+} \rightarrow e^{+} +
\nu_{\mu}} & \simeq &  -\frac{i \,g }{4\sqrt{2}(2 \pi)^{3/2}}
 \; \frac{\varepsilon _{{\bf p},\mu,\lambda}}{\sqrt{2 E^{W}_{p}}}\; \delta^{3}({\bf p}-{\bf
q}-{\bf k})\; \sin 2\theta  \; \De t \,\frac{\Delta m}{2 k}
\\ & \times &
\, \overline{v}_{-{\bf k},1}^{r}
\; \gamma^{\mu}(1 - \gamma^{5})\; v_{{\bf q},e}^{s}\,.
\eea

 Eqs.(\ref{amplitude-nuePontec2})
and (\ref{amplitude-numuPontec5}) can be combined to give the  branching ratio
\bea\label{branch}
\frac{\Gamma(W^{+} \rightarrow e^{+} + \nu_{\mu})}{\Gamma(W^{+} \rightarrow e^{+} + \nu_{e})}\;
\sim \; \sin^{2}2\theta \; \frac{(\Delta m)^2}{4 k^2}\,.
\eea
This result clearly shows that the use of Pontecorvo flavor states leads to a violation of the
lepton charge in the production vertex.
The result (\ref{branch}) is derived in the relativistic limit; however
the lepton charge violation effect  is more significant in the non-relativistic
region
(see Eqs.(\ref{amplitude-nuePontec}) and (\ref{amplitude-numuPontec3})).

In the above treatment, we have not considered explicitly the $W^{+}$ decay width
$\Gamma$. This should be taken into account when comparing our results with the ones
of Ref.\cite{Nishi:2008sc}.
However, the fact that the amplitude $A_{W^{+} \rightarrow e^{+} +
\nu_{\mu}}$ calculated with the exact flavor states
vanishes is independent of the inclusion of the decay width in the calculation.

\section{Conclusions \label{S5}}

In this paper, we have analyzed the amplitudes of the weak
interaction processes where flavor neutrinos are generated. We have
done explicit computations at tree level for the processes $W^{+}
\rightarrow e^{+} + \nu_{e}$ and $ W^{+} \rightarrow e^{+} +
\nu_{\mu}$ using
 the exact flavor states and
the Pontecorvo states.
We have  considered the above amplitudes in the
short time limit, i.e. at very small distances from the production
vertex. In this case, we found that the use of the exact flavor states in the
computations leads to
consistent results, whereas the Pontecorvo states  yield a violation of the
lepton charge in the vertex. Consistency with the SM phenomenology
is thus attained only for the  QFT exact flavor states.

In order to better understand the results presented above, we  observe that the amplitudes in
the short time limit
give information on the decay process very close to the vertex.
Thus, one can associate  a wavefunction, say $u_{{\bf k},\nu_e}^{r}$,
with the electron neutrino   in the amplitudes given by Eqs.(\ref{amplitude-nue6}) and
(\ref{amplitude-nuePontec}).
In the case of exact flavor states, the amplitude given by Eq.(\ref{amplitude-nue6}) suggests
that
$u_{{\bf k},\nu _{e}}^{r} =u_{{\bf k},1}^{r}$, i.e.
the  wavefunction for $\nu_e$ is the same as the one for $\nu_1$, with
$u_{{\bf k},1}^{r \dag} u_{{\bf k},1}^{r}=1$.
On the other hand, in the case of Pontecorvo states, the amplitude given by
Eq.(\ref{amplitude-nuePontec}) leads to  the identification:
\bea\lab{wfPont}
u_{{\bf k},\nu _{e}}^{r} = \cos^{2}\theta \;{u}_{{\bf k},1}^{r} + \sin^{2}\theta
\;{u}_{{\bf k},2}^{r}\,.
\eea
Such a wavefunction, however, is not normalized properly as one can easily see:
 \bea
u_{{\bf k},\nu _{e}}^{r \dag} u_{{\bf k},\nu _{e}}^{r} =
\cos^{4}\theta \;+ \sin^{4}\theta \;+ 2 \, \sin^{2}\theta\,
\cos^{2}\theta \,|U_{\bf k}|\,,
\eea
where we have used  Eq.(\ref{Uk2}). Since $|U_{\bf k}| < 1$ for $m_1 \neq m_2$, the
above wavefunction is not normalized.

Note also that the amplitude
Eq.(\ref{amplitude-numuPontec3}) contains the combination  $u_{{\bf
k},\De \nu_e}^{r}\equiv(u_{{\bf k},2}^r -u_{{\bf k},1}^{r})
\sin\theta \cos\theta $, which  is also not normalized:
\bea
u_{{\bf k},\De\nu_e}^{r \dag} u_{{\bf k},\De\nu_e}^{r} =  2\,\sin^{2}\theta
\cos^{2}\theta\,\Big(1- \,|U_{\bf k}| \Big ).
\eea
This is just the missing piece necessary for the normalization of $u_{{\bf k},\nu _{e}}^{r}$
in Eq.(\ref{wfPont}):
\bea
 u_{{\bf k},\De \nu_e}^{r \dag} u_{{\bf k},\De \nu_e}^{r} \, +\,u_{{\bf k},\nu _{e}}^{r \dag}
u_{{\bf k},\nu _{e}}^{r} \,=\,1.
\eea

In conclusion, a violation of lepton charge in the production vertex
is due to the incorrect treatment of the flavor neutrino states.
Defining them as the eigenstates of flavor charges \cite{BHV98,currents},  results consistent with
 Standard Model are found.

\section*{Acknowledgements}

We thank C. Giunti for stimulating discussions. Support from
INFN is also acknowledged. The work of C.-R.Ji was supported
in part by the U.S. Department of Energy(No. DE-FG02-03ER41260).

\appendix
\section{The vacuum structure for fermion mixing}

We briefly summarize the QFT formalism of the neutrino mixing.  For
a detailed review see \cite{Capolupo:2004av}. The mixing
transformations are
 \bea\label{mix} \nu _{e}(x) &=&\cos \theta \,\nu_1(x) \,+ \,\sin \theta \,\nu _{2}(x)
\\[2mm]
 \non \nu _{\mu }(x)
&=&-\sin \theta\, \nu _{1}(x)\, +\,\cos \theta \,\nu _{2}(x)\, ,
\eea
where
$\nu_{e}(x)$ and $\nu_{\mu}(x)$ are the Dirac neutrino fields with
definite flavors. Here, $\nu_{1}(x)$ and $\nu_{2}(x)$ are the free
 neutrino
fields with definite masses $m_{1}$ and $m_{2}$, respectively. The
fields $\nu_{1}(x)$ and $\nu_{2}(x)$ can be written as
\bea\label{freefi}
 \nu _{i}(x)=\frac{1}{\sqrt{V}}{\sum_{{\bf k} ,
r}} \left[ u^{r}_{{\bf k},i}\, \al^{r}_{{\bf k},i}(t) +
v^{r}_{-{\bf k},i}\, \bt^{r\dag}_{-{\bf k},i}(t) \ri] e^{i {\bf
k}\cdot{\bf x}},\qquad \qquad i=1,2
\eea
 with $ \al_{{\bf
k},i}^{r}(t)=\al_{{\bf k},i}^{r}\, e^{-i\omega _{k,i}t}$, $
\bt_{{\bf k},i}^{r\dag}(t) = \bt_{{\bf k},i}^{r\dag}\,
e^{i\omega_{k,i}t},$ and $ \omega _{k,i}=\sqrt{{\bf k}^{2} +
m_{i}^{2}}.$ The operator $\alpha ^{r}_{{\bf k},i}$ and $ \beta
^{r }_{{\bf k},i}$, $ i=1,2 \;, \;r=1,2$ are the annihilator
operators for the vacuum state
$|0\rangle_{m}\equiv|0\rangle_{1}\otimes |0\rangle_{2}$: $\alpha
^{r}_{{\bf k},i}|0\rangle_{m}= \beta ^{r }_{{\bf
k},i}|0\rangle_{m}=0$.
 The anticommutation relations are:
$\left\{ \nu _{i}^{\alpha }(x),\nu _{j}^{\beta \dagger
}(y)\right\} _{t=t^{\prime }}=\delta ^{3}({\bf x-y})\delta
_{\alpha \beta } \delta _{ij},$ with $\alpha ,\beta = 1,...4,$
and
$\left\{ \alpha _{{\bf k},i}^{r},\alpha _{{\bf q},j}^{s\dagger
}\right\} =\delta _{{\bf kq}}\delta _{rs}\delta _{ij};$ $\left\{
\beta _{{\bf k},{i}}^{r},\beta _{{\bf q,}{j}}^{s\dagger }\right\}
=\delta _{{\bf kq}}\delta _{rs}\delta _{ij},$ with $i,j=1,2.$ All
other anticommutators vanish. The orthonormality and
completeness relations are given by $u_{{\bf k},i}^{r\dagger }u_{{\bf
k},i}^{s} = v_{{\bf k},i}^{r\dagger }v_{{\bf k},i}^{s} = \delta
_{rs},\; $ $u_{{\bf k},i}^{r\dagger }v_{-{\bf k},i}^{s} = v_{-{\bf
k} ,i}^{r\dagger }u_{{\bf k},i}^{s} = 0,\;$ and $\sum_{r}(u_{{\bf
k},i}^{r}u_{{\bf k},i}^{r\dagger }+v_{-{\bf k},i}^{r}v_{-{\bf
k},i}^{r\dagger }) = 1.$

The generator of the mixing transformations
 is given by  \cite{BV95}:
\bea\label{generator12}
 G_{\bf \te}(t) = \exp\left[\theta \int
d^{3}{\bf x} \left(\nu_{1}^{\dag}(x) \nu_{2}(x) -
\nu_{2}^{\dag}(x) \nu_{1}(x) \right)\right]\;
\eea
and $ \nu_{\sigma}^{\alpha}(x) = G^{-1}_{\bf
\te}(t)\;
\nu_{i}^{\alpha}(x)\; G_{\bf \te}(t)$ for $(\sigma,i)=(e,1)$ and $(\mu,2)$.
At finite volume, this is a unitary operator, $G^{-1}_{\bf
\te}(t)=G_{\bf -\te}(t)=G^{\dag}_{\bf \te}(t)$, preserving the
canonical anticommutation relations. The generator $G^{-1}_{\bf
\te}(t)$ maps the Hilbert space for free fields ${\cal H}_{m}$ to
the Hilbert space for mixed fields ${\cal H}_{f}$: $
G^{-1}_{\bf \te}(t): {\cal H}_{m} \mapsto {\cal H}_{f}.$ In
particular, the flavor vacuum is given by
$
 |0(t) \rangle_{f} = G^{-1}_{\bf \te}(t)\;
|0 \rangle_{m}\; $ at finite
volume $V$. We denote by
$|0 \rangle_{f}$  the flavor vacuum  at $t=0$.
In the infinite volume limit, the flavor  and the mass vacua
are unitarily
inequivalent \cite{BV95,JM01}. Similarly, flavor vacua at different times are orthogonal \cite{Blasone:2005ae}.
The flavor fields are written as:
\begin{eqnarray}\label{flavorfield}
\nu _{\sigma}({\bf x},t) &=&\frac{1}{\sqrt{V}}{\sum_{{\bf k},r} }
e^{i{\bf k.x}}\left[ u_{{\bf k},i}^{r}\,  \alpha _{{\bf
k},\nu_{\sigma}}^{r}(t) + v_{-{\bf k},i}^{r} \, \beta _{-{\bf
k},\nu_{\sigma}}^{r\dagger }(t)\right]\,,
\end{eqnarray}
with $(\sigma,i)=(e,1),(\mu,2).$
 The flavor annihilation operators are  \cite{BV95}:
\bea \non \alpha^{r}_{{\bf k},\nu_{e}}(t)&=&\cos\theta\;\alpha^{r}_{{\bf
k},1}(t)\;+\;\sin\theta\;\sum_{s}\left[u^{r\dag}_{{\bf k},1}
u^{s}_{{\bf k},2}\; \alpha^{s}_{{\bf k},2}(t)\;+\; u^{r\dag}_{{\bf
k},1} v^{s}_{-{\bf k},2}\; \beta^{s\dag}_{-{\bf k},2}(t)\right]
\\ \non
\beta^{r}_{-{\bf k},\nu_{e}}(t)&=&\cos\theta\;\beta^{r}_{-{\bf k},1}(t)\;+
\;\sin\theta\;\sum_{s}\left[v^{s\dag}_{-{\bf k},2} v^{r}_{-{\bf
k},1}\; \beta^{s}_{-{\bf k},2}(t)\;+\; u^{s\dag}_{{\bf k},2}
v^{r}_{-{\bf k},1}\; \alpha^{s\dag}_{{\bf k},2}(t)\right]
\label{annih1} \eea
and similar expressions hold for muonic neutrinos.
In the reference frame where ${\bf k}=(0,0,|{\bf k}|)$, we have
\bea\label{annih1K00}
\alpha^{r}_{{\bf k},\nu_{e}}(t)&=&\cos\theta\;\alpha^{r}_{{\bf
k},1}(t)\;+\;\sin\theta\;\left( |U_{{\bf k}}|\; \alpha^{r}_{{\bf
k},2}(t)\;+\;\epsilon^{r}\; |V_{{\bf k}}|\; \beta^{r\dag}_{-{\bf
k},2}(t)\right),
\\
\beta^{r}_{-{\bf k},\nu_{e}}(t)&=&\cos\theta\;\beta^{r}_{-{\bf
k},1}(t)\;+\;\sin\theta\;\left( |U_{{\bf k}}|\; \beta^{r}_{-{\bf
k},2}(t)\;-\;\epsilon^{r}\; |V_{{\bf k}}|\; \alpha^{r\dag}_{{\bf
k},2}(t)\right),
\eea
and similar ones for $\alpha^{r}_{{\bf k},\nu_{\mu}}$
 and $\beta^{r}_{-{\bf k},\nu_{\mu}}$.
 In Eq.(\ref{annih1K00}), $\epsilon^{r}=(-1)^{r}$ and
\bea\label{Vk2} \non |V_{{\bf k}}| & \equiv & \epsilon^{r}\; u^{r\dag}_{{\bf
k},1} v^{r}_{-{\bf k},2} = -\epsilon^{r}\; u^{r\dag}_{{\bf k},2}
v^{r}_{-{\bf k},1}\,
=\,\frac{ (\om_{k,1}+m_{1}) - (\om_{k,2}+m_{2})}{2
\sqrt{\om_{k,1}\om_{k,2}(\om_{k,1}+m_{1})(\om_{k,2}+m_{2})}}\, |{\bf k}| \,;
\\
\\ \label{Uk2}
 |U_{{\bf k}}| & \equiv & u^{r\dag}_{{\bf k},i}
u^{r}_{{\bf k},j} = v^{r\dag}_{-{\bf k},i} v^{r}_{-{\bf k},j} \,
=\,\frac{|{\bf k}|^{2} +(\om_{k,1}+m_{1})(\om_{k,2}+m_{2})}{2
\sqrt{\om_{k,1}\om_{k,2}(\om_{k,1}+m_{1})(\om_{k,2}+m_{2})}}\,,
\eea
with $i,j = 1,2,  i \neq j.$
We have:
%
$|U_{{\bf
k}}|^{2}+|V_{{\bf k}}|^{2}=1$.
Note that the following relations hold:
\bea \label{reu2} \overline{u}_{{\bf k},1}^{r}\;\sum_{s}
u^{r\dag}_{{\bf k},1} u^{s}_{{\bf k},2} + \overline{v}_{-{\bf
k},1}^{r} \;\sum_{s} v^{r\dag}_{-{\bf k},1} u^{s}_{{\bf k},2} & =
& \overline{u}_{{\bf k},2}^{r}\,,
\\
\label{rev2} \overline{u}_{{\bf k},1}^{r}\;\sum_{s}
u^{r\dag}_{{\bf k},1} v^{s}_{-{\bf k},2}
 +  \overline{v}_{-{\bf k},1}^{r}
\sum_{s} v^{r\dag}_{-{\bf k},1} v^{s}_{-{\bf k},2} & = &
\overline{v}_{-{\bf k},2}^{r}\,. \eea
In the reference frame where ${\bf k}=(0,0,|{\bf k}|)$, they become
\bea \label{realzu2} \overline{u}_{{\bf k},1}^{r}\;|U_{\bf k}| -
\varepsilon^{r}\, \overline{v}_{-{\bf k},1}^{r}|V_{\bf k}| & = &
\overline{u}_{{\bf k},2}^{r}
\\[2mm]
\label{realzv2} \overline{u}_{{\bf k},1}^{r}\;|V_{\bf k}| +
\varepsilon^{r}\, \overline{v}_{-{\bf k},1}^{r}|U_{\bf k}| & = &
\varepsilon^{r}\, \overline{v}_{-{\bf k},2}^{r}\;.
\eea
Moreover, we have
\bea \label{realzu3} \overline{u}_{{\bf k},2}^{r}\;|U_{\bf k}| +
\varepsilon^{r}\, \overline{v}_{-{\bf k},2}^{r}|V_{\bf k}| & = &
\overline{u}_{{\bf k},1}^{r}\;.
\eea

\section{Comment on the amplitudes in the long time limit}

Let us now comment on  the amplitudes for the decay processes (\ref{We-decay}) and
(\ref{Wmu-decay}) computed in the long time limit as done in Ref.\cite{Li:2006qt}.
There it was argued that the non zero result in such a limit for the amplitude Eq.(\ref{Wmu-decay})
implies a flavor violation. We point out, however, that such a ``problem'' arises also with Pontecorvo states.
Indeed, considering $x^{0}_{in} \rightarrow -\infty$ and $x^{0}_{out}
\rightarrow +\infty$, the non-vanishing amplitude $A^{P}_{W^{+} \rightarrow e^{+} + \nu_{\mu}} $ is obtained from 
 Eq.(\ref{amplitude-numuPonte1}):
\bea\label{amplitude-numuPonteINF}\non
  A^{P}_{W^{+} \rightarrow e^{+} + \nu_{\mu}} & = & \frac{i \,g }{2\sqrt{2}(2 \pi)^{1/2}}
 \; \frac{\varepsilon _{{\bf p},\mu,\lambda}}{\sqrt{2 E^{W}_{p}}}\; \sin \theta\; \cos
\theta\,\delta^{3}({\bf p}-{\bf q}-{\bf k})\,
\\ \non & \times &
 \Big [e^{- i  \omega_{k,2} x^{0}_{out}}\;
\overline{u}_{{\bf k},2}^{r}\; \gamma^{\mu}\; (1 - \gamma^{5})\;
v_{{\bf q},e}^{s} \; \delta (E^{W}_{p} - E^{e}_{q} - \omega_{k,2})
\; \\ & - &
 \;e^{- i  \omega_{k,1} x^{0}_{out}}\; \overline{u}_{{\bf k},1}^{r}\;
  \gamma^{\mu}\; (1 - \gamma^{5})\; v_{{\bf q},e}^{s} \;
 \; \delta (E^{W}_{p} - E^{e}_{q} - \omega_{k,1})\Big ] \;.
\eea

In a similar way, the amplitude $A^{P}_{W^{+} \rightarrow e^{+} + \nu_{e}}$
Eq.(\ref{amplitude-nuePonte1}) becomes
\bea
\label{amplitude-nuePonteINF}
 A^{P}_{W^{+} \rightarrow e^{+} + \nu_{e}} & = &  \frac{i \,g }{2\sqrt{2}(2 \pi)^{1/2}}
 \; \frac{\varepsilon _{{\bf p},\mu,\lambda}}{\sqrt{2 E^{W}_{p}}}\;\delta^{3}({\bf p}-{\bf q}-{\bf k})\;
\\ \non & \times &
 \Big [\cos^{2}\theta \;e^{- i  \omega_{k,1} x^{0}_{out}}\;
\overline{u}_{{\bf k},1}^{r}\; \gamma^{\mu}(1 - \gamma^{5})\;
v_{{\bf q},e}^{s}\, \delta(E^{W}_{p} - E^{e}_{q} - \omega_{k,1})
 \\ \non & + &
  \sin^{2}\theta \;e^{- i  \omega_{k,2} x^{0}_{out}}\; \overline{u}_{{\bf k},2}^{r}\;
  \gamma^{\mu}(1 - \gamma^{5})\; v_{{\bf q},e}^{s} \; \delta (E^{W}_{p} - E^{e}_{q} - \omega_{k,2})
\Big ]\;.
\eea

For the exact flavor states, one obtains from Eq. (\ref{amplit2}) (and Eq. (\ref{amplitude2expl2}))
results which reproduce Eq. (\ref{amplitude-nuePonteINF}) (and Eq. (\ref{amplitude-numuPonteINF})) in the relativistic limit.

As already observed, the mixed neutrinos cannot be considered as asymptotic fields.
Considering then the long time limit amounts to average over the flavor oscillations.
Thus it is not surprising that the amplitude $A_{W^{+} \rightarrow e^{+} + \nu_{\mu}} $
gives a non zero result. In the long time limit the energy conservation is  made explicit
by the presence of the delta functions.

For the case of exact flavor states, the obtained results reproduce Eqs.(3.9)
and (3.13) of Ref. \cite{Li:2006qt}. In such a case,
terms due to the neutrino condensate are also present and are proportional
to the $|V_{\bf k}|$ function. We point out that one should not be misled (as in
Ref.\cite{Li:2006qt}) by the sign of the corresponding energies in the
 delta functions, since the negative  $\omega_{k,2}$,
 appearing in Eqs.(3.9)
and (3.13) of Ref. \cite{Li:2006qt}, is associated to ``hole'' contributions
 in the flavor vacuum condensate. Contrary to
the claim of the authors of \cite{Li:2006qt}, there is nothing
paradoxical or wrong  in these signs.

\end{document}